\documentstyle{mn}

\title{9 Aurigae: strong evidence for non-radial pulsations}
\author[K. Krisciunas et al.]
       {K. Krisciunas$^1$, R. F. Griffin$^2$, E. F. Guinan$^3$,
        K. D. Luedeke$^4$, and G. P. McCook$^3$  \\
      $^1$Joint Astronomy Centre, 660 N. A'oh\={o}k\={u} Place, University
Park, Hilo, HI 96720, USA \\
      $^2$The Observatories, Madingley Road, Cambridge CB3 0HA, England, UK \\
      $^3$Villanova University, Astronomy Department, Villanova, PA 19085,
USA \\
      $^4$9624 Giddings Avenue NE, Albuquerque, NM 97109, USA}
\date{Accepted .................................
      Received .................................;
      in original form  .................................}

\begin{document}

\maketitle

\begin{abstract}
We present further photometric observations of the unusual F0 V
star 9 Aurigae and present evidence that this star's radial velocity,
spectroscopic line widths and line depths are also variable with the
same frequencies as the photometric data ($f_1 \approx 0.795$ and
$f_2 \approx 0.345$ d$^{-1}$).  The phases of these sinusoids are stable
over time scales of longer than one year, though the amplitudes can vary,
making the prediction of photometric behavior impossible.
Given that a variety of other
explanations have already been discounted (e.g. interactions with a
close companion, the existence of a lumpy, orbiting ring of dust, or star
spots) and that these variations occur
on time scales an order of magnitude slower than the fundamental radial
pulsation period, we have very strong evidence that 9 Aurigae exhibits
non-radial {\it g}-mode pulsations. Since the power spectrum of the
radial velocity data shows frequency $f_2$ but does not clearly show $f_1$,
the present data suggest
that $f_2$ is associated with a low degree spherical harmonic ($\ell$ =
1 or 2), while $f_1$ is associated with a higher degree harmonic.
9 Aurigae, along with such stars as $\gamma$ Doradus,
HD 224638, HD 224945, and HD 164615, appear to constitute a new
class of pulsating variables.  These stars are to be found at or beyond the
cool edge of the Cepheid instability strip in the HR Diagram.  Prior to
this, only much hotter stars have been shown to exhibit non-radial
{\it g}-modes.

\end{abstract}

\section{Introduction}

Most spectroscopically normal, single F-type dwarf stars are very constant
in brightness.  Recently, a number of F-type stars on, or just above, the
main sequence in the Hertzsprung-Russell (HR) Diagram have been been shown
to be photometrically variable, up to 0.1 magnitude on time scales of several
hours to tens of hours.

The first of these to be identified was the
bright southern star, $\gamma$ Doradus (Cousins \& Warren 1963).  It was
most recently studied by Balona {\em et al.} (1994a) and by
Balona, Krisciunas \& Cousins (1994b).  A number of
other stars of similar spectral type and luminosity class have been
independently identified to be variable in brightness.  This includes
HD 164615 (Abt, Bollinger \& Burke 1983), four peculiar variables in the
open cluster NGC 2516 (Antonello \& Mantegazza 1986), HD 96008 (Lampens 1987;
Matthews 1990), 9 Aurigae (Krisciunas and Guinan 1990; Krisciunas {\it et al.}
1993), HD 111829 (Mantegazza, Poretti \& Antonello 1991; note correct HD
number), and HD 224638 and HD 224945
(Mantegazza \& Poretti 1991; Mantegazza, Poretti \& Zerbi 1994).  Because
these stars have such similar spectral types and absolute magnitudes,
hence temperatures and densities, it is sensible to study their similarities
under the assumption that they constitute a new group of variables.

Antonello \& Mantegazza (1986) first suggested that the slowly varying F stars
in NGC 2516 might be undergoing non-radial pulsations.  However, they
felt that such a notion would be difficult to verify because the expected
radial velocity and color variations would ``probably [be] beyond the
present precision of measurements.''  They suggested star spots and
ellipsoidally shaped stars as other possibilities.
Waelkens (1991) independently suggested that one of the stars mentioned
above, HD 96008, was undergoing non-radial pulsations.  However,
Mantegazza \& Poretti (1994) have recently shown that HD 96008 exhibits
smooth radial velocity variations with a period equal to twice the
photometric period.  It is almost certain that HD 96008 is an ellipsoidal
star with a close companion (likely to be an M5 dwarf star).

The reason for the suggestion that one or more of these stars is
undergoing non-radial pulsations is that a star with the density of an F
dwarf has a fundamental radial pulsation period of
1-3 hours, and non-radial gravity modes give
rise to variability on longer time scales.  If an early F
dwarf has a period of 18 hours ($\gamma$ Dor), or more than one day
(9 Aur), it could be variable as a result of non-radial gravity modes.

The proof for this is two-fold. First, one looks for the
{\em absence} of certain kinds of evidence.

A zeroth order consideration is whether it  can be proven that 9 Aur is
{\em not} a $\delta$ Scuti star.  This involves looking for evidence of
variations on time scales of $\approx$ 0.5 to 5 hours.  Data sufficient
for this test are to be found in IAU files 238 and 244 of Unpublished
Observations of Variable Stars (see Breger, Jaschek \& Dubois 1990).  This
issue was particularly addressed by Krisciunas {\em et al.} (1991).  On one
occasion Abt obtained 28 radial velocities over 2.0 hours.  Guinan twice
obtained more than 50 differential magnitudes in 1.5 hours;  Skillman
one night obtained 35 photometric points in 5.6 hours; Luedeke obtained
27 points in 4.1 hours; Ohshima obtained 63 points over 5.0 hours; and
Krisciunas obtained 40 points over 5.2 hours.  From light curves and
power spectra of such data obtained on single nights or several nights in
a row there is no evidence of variations with frequencies of 5 to 44 d$^{-1}$,
something one would expect {\em if} 9 Aur were a $\delta$ Scuti star.
(See Fig. 1 for an example.)  We note, however, that variations of low
amplitude ($<$ 2 millimagnitudes) in the $\delta$ Scuti regime of
frequencies are not ruled out by the data.

 \begin{figure}
 \vspace{70mm}
 \caption{Power spectrum of the V-band photometry of 1992 February 7/8 to
  February 9/10, based on data by Krisciunas, Ohkura, and Akazawa and found
  in IAU file 244. This and power spectra of other data referenced in the
  text demonstrate that the photometric variations of 9 Aurigae occur
  on time scales much slower than $\delta$ Scuti stars.}
 \end{figure}

If there is no evidence for the existence of a close, possibly
interacting companion, one can rule out the ellipsoidal star
hypothesis and the notion that there is mass transfer going
on.  From the light curve one can tell if the star is an
eclipsing binary.  If a star's radial velocity is constant to within
a few km s$^{-1}$, the star is not likely to be a spectroscopic binary.

One of the ideas considered by a number of authors is the existence
of extensive star spot areas, which give rise to small amplitude
variability as the star rotates.  (This is the mechanism for the
variability of BY Dra stars, which are K and M dwarfs.)  But if a dwarf star
has a spectral type earlier than about F5 to F7 and shows no
evidence for an active chromosphere, then it will likely not
have extensive spot areas (Giampapa \& Rosner 1984).
On the basis of the two closely spaced periods of $\gamma$ Dor and the
assumption that they could be explained by differential rotation on a
spotted star, Balona {\em et al.} (1994b)
attempted to model that star with a star spot model.  They
concluded, however, that the star spot model is not very plausible.
For a star with non-commensurate periods of photometric variation the star spot
model does not work at all.

If a star has a single, clearly defined photometric period (here
assumed to be equal to the rotational period), under the assumption that
we know the radius of the star from its location in the HR Diagram, we can
use the projected rotational rate ({\it v} sin {\it i}) to
derive the inclination angle {\it i}.  If it follows that the star is
viewed nearly pole on ({\it i} near 0), there is essentially no ``horizon''
beyond which any presumed star spots could disappear.  In a previous paper
(Krisciunas {\em et al.} 1993, hereafter referred to as Paper I)
we found periods of 1.277 and 2.725 days for 9 Aur from a multi-longitude
photometry campaign.  Given the star's size of 1.64 $R_{\odot}$
(Mantegazza {\em et al.} 1994) and {\it v} sin {\it i} = 17.80
km s$^{-1}$ (see below), we can calculate the implied inclination angle
if we know the rotational period of the star.  (While Mantegazza {\em et al.}
do not give an uncertainty for the size of the star, we shall optimistically
adopt $\pm 0.2 R_{\odot}$.)  If the rotation period is 1.277 days,
the implied inclination angle is 16 $\pm$ 2 degrees.  If the
longer photometric period is the rotational period, the inclination angle
would be 36 $\pm$ 5 degrees, which in fact {\em would} allow star spots to be
a possible explanation, since some known spotted stars have inclination
angles on this order.  Given the reasons outlined above and the
absence of chromospheric activity in our slowly varying F stars, it seems
unlikely that star spots are the physical cause of their variability.

The other kind of proof is more in the realm of
confirmation.  If a star is undergoing low degree non-radial pulsations
($\ell \leq 2$) one might measure radial velocity variations of a few
km s$^{-1}$.  For these and especially higher degree harmonics
one might observe line profile variations.
Examples of the type of line profile variations one might observe (for
$\ell$ = 1 to 10 {\em sectoral} modes) are given by Vogt \& Penrod (1983).
For general information on non-radial pulsations, see Cox (1980)
and  Unno {\em et al.} (1989).

Balona {\em et al.} (1994a) report line profile variations for
$\gamma$ Doradus, and Mantegazza {\em et al.} (1994) show
that HD 224638 can exhibit minor changes in line profile.  In
the case of these two stars the line profile variations are
based on 9 and 3 spectra, respectively.  Neither star seems
to exhibit radial velocity variations.

In Paper I we sought a variety of evidence for the cause of
the variability of the bright northern star 9 Aurigae.  From
UKIRT, IRAS, IUE, and speckle data, it showed no evidence for
a close companion or an orbiting lumpy ring of dust.  (It has
a companion about 5 arcsec distant, which corresponds to
about 100 AU.  From infrared photometry this is most likely
an M2 dwarf star.  9 Aur A and B are too far apart to
interact in any significant way, and the fainter star, being
7 magnitudes fainter in V, is much too faint to affect the
optical photometry.) \footnote [1] {Hereafter, when we say 9 Aur
we mean the primary, 9 Aur A.}

In our search for an explanation for 9 Aurigae's
variability one of us (RFG) obtained 22 radial velocities
with the Coravel radial velocity spectrometer at Haute
Provence Observatory.  These were obtained on 1992 April 23
and from 1993 February 10 to March 24 on a total of 14 nights
and showed a range of about 7 km s$^{-1}$.  With an internal error
per measurement of about 0.6 km s$^{-1}$, the variations seemed
significant.  Furthermore, a period of just under 3 days was
indicated.  What was clearly needed was a more homogeneous
data set -- one measurement per hour for as many hours and
nights as possible.  Given the type of data Coravel produces,
this would also allow us to see if the radial velocities,
line widths, fractional line depths, and line profiles vary in a
smooth way from hour to hour and from night to night.

In this paper we report extensive data runs of
photometry and radial velocities of 9 Aur.  Given the absence
of certain types of evidence (from Paper I) and the data
presented in this paper, we believe we have very strong
evidence that 9 Aur is exhibiting non-radial pulsations.
Given the time scales involved, they would have to be gravity
modes.

\section{Observations}

In Table 1 we give a summary of the V-band and B-band
photometry.  The individual data values can be obtained by
requesting IAU file 285E of Unpublished Observations of
Variable Stars.

\begin{table}
\caption{Summary of 9 Aur photometry contained in IAU file 285E.  Given is
the telescope aperture,
the number of nights N on which observations were made, and the number
of V-band and B-band observations with respect to BS 1561.  Each of
Guinan's points for the 1993/4 season represents the mean of 6 to 10
individual differential measures.}
\begin{tabular}{lrcrrr}\hline

$Observer$ &  $Season$ & $Aper \hspace{1 mm} (cm)$ &  $N$ & $n_V$ & $n_B$ \\
\hline

Guinan     & 1992/3 &  76   &  13  &   177  &  177 \\
Krisciunas & 1992/3 &  15   &   3  &    13  &      \\
Guinan     & 1993/4 &  76   &  59  &   115  &  115 \\
Luedeke    & 1993/4 &  20   &  11  &    86  &      \\
Krisciunas & 1993/4 &  15   &   2  &    21  &      \\

\hline
\end{tabular}
\end{table}

Guinan's data, reduced by McCook (hereafter called the Guinan and McCook
data), were obtained at Mt. Hopkins, Arizona,
with two 76-cm Automatic Photoelectric Telescopes (APTs).  One APT is
operated by the Four College Consortium (FCC), the other by Fairborn
Observatory.  These two telescopes have photometers with matching
photomultiplier tubes and filters.
The APT observations were made using a 2.5 magnitude neutral density
filter to reduce the stars' count rates to reasonable levels.
Luedeke's data were obtained in Albuquerque, New Mexico.
Krisciunas' data were obtained at the 2800-m elevation of
Mauna Kea, Hawaii.  As in Paper I, the photometry of 9 Aur
was obtained by means of differential measures with respect
to BS 1561 (= HD 31134; V = +5.78; Sp = A2 V).
Krisciunas used BS 1568 as a check star, while
Luedeke primarily used BS 1668 as a check star.  From
over 300 differential measures with respect to BS 1568 and BS
1668 obtained over several years, we find no evidence that
our principal comparison star, BS 1561, is variable in any
way.  Hence any variations of 9 Aur vs. BS 1561 we attribute
solely to 9 Aur.  These comparison star vs. check star
observations also tell us how accurate an individual
differential measurement is.  For the Guinan and McCook data it is
$\pm 10$ millimagnitudes (mmag) or better.  For the Luedeke data the
corresponding value is $\pm 12$ mmag, and for the most
recent Krisciunas data it is  $\pm 20$ mmag.  Guinan's
1993/4 data are averages of 6 to 10 differential measures per
point, implying internal errors of 3 to 5 mmag per point.

    The radial velocities were obtained by RFG with the Haute Provence
Coravel (Baranne, Mayor \& Poncet 1979).   A scanning range of 70 km s$^{-1}$,
centred on zero heliocentric velocity, was normally used and was wide enough
to encompass most (usually all) of the width of the cross-correlation dip.
The normal integration time was 5 minutes, which in most cases was long enough
to give an ample signal but was needed in order to reduce 'seeing noise'.
Such noise, arising from the fluctuations in the amount of light
passing the entrance slit of the Coravel spectrometer, can be objectionable
in short integrations, which average too few of the 5-Hz scans; the effects
are particularly noticeable in traces, such as are given by 9 Aur, exhibiting
wide and shallow dips.

During one particular observing run (a long run
near to opposition of 9 Aur) an intensive series of measurements was made, at
rather strictly timed hourly intervals whenever weather and other circumstances
permitted.   On one night as many as 14 consecutive hourly observations were
obtained.

\section{Analysis of photometry}

Since Guinan and McCook's latest data represent a number of differential
measures per point, some averaging was necessary for the
Krisciunas and Luedeke data, so that the data would have more
comparable weighting.  Given that the minimum number of
differential measures per night was 3, we averaged the
Krisciunas and Luedeke data by groups of 3.

 \begin{figure}
 \vspace{70mm}
 \caption{Power spectrum of the V-band photometry of 1993 September 3/4 to
  1994 February 5/6.}
 \end{figure}

In Fig. 2 we show the power spectrum of the 1993/4
data, using the Lomb-Scargle algorithm as presented by Press
\& Teukolsky (1988).  It is cleaner than the power spectrum
of the data of early 1992 presented in Paper I, which was
based on only an 11 day observing campaign.  In Paper I we found principal
frequencies of $\approx$ 0.783 and $\approx$ 0.367 d$^{-1}$.  The 1993/4
data give $f_1$ = 0.79475 and $f_2$ = 0.345684 d$^{-1}$.
If we combine the last three seasons of data, we get $f_1$ = 0.7948 and
$f_2$ = 0.3456.  Because of gaps in the photometry, particularly
from season to season, and noting the relative complexity of
the power spectra of different combinations of data,
we decided that the most accurate working frequencies were obtained from
the latest season. In this paper since we are primarily interested in the
photometry and radial velocities of the latest season, using the
frequencies derived from the latest data seems sensible.  In section 5
below we suggest slight revisions to these periods for longer term studies.

Once we decided on the working frequencies,
software obtained from Luis Balona allowed us to fit the data according
to the following functional form:\\

$$\Delta V = A_1\cos{2\pi(f_1t + \phi_1)} + A_2\cos{2\pi(f_2t + \phi_2)}
 + ... $$ \\

\noindent
giving the amplitudes $A_i$ and the phases $\phi_i$ of the
sinusoids given the known frequencies $f_i$.  In Table 2 we
give the least-squares results of the 1993/4 photometry.

\begin{table}
\caption{Amplitudes ($A_i$, in mmag) and phases ($\phi_i$) of Fourier fit
to the 1993/4 V-band data using epoch JD 2449000.0 and frequencies
$f_1 = 0.79475$ and $f_2 = 0.345684$ d$^{-1}$.  The standard deviation
of an individual data point in the two frequency fit is $\pm$ 12.1 mmag.}
\begin{tabular}{crr}\hline
$i$ & \multicolumn{1}{c}{$A_i$} &  \multicolumn{1}{c}{$\phi_i$} \\
\hline
  $1$  &  $13.8 \pm 1.4$ & $-0.114 \pm 0.016$ \\
  $2$  &  $12.2 \pm 1.4$ & $-0.481 \pm 0.018$ \\
\hline
\end{tabular}
\end{table}

 \begin{figure}
 \vspace{140mm}
 \caption{(a) Power spectrum of the V-band photometry, with $f_2$ sinusoid
  subtracted out.  The frequency $f_1$ and its one day alias are clearly
  present. (b) Power spectrum of the V-band photometry, with $f_1$ sinusoid
  subtracted out. The frequency $f_2$ and its one day alias are clearly
  present.}
 \end{figure}

Fig. 3{\em a} shows the corresponding power spectrum if the
sinusoid associated with frequency $f_2$ is subtracted from
the data set.  Similarly, Fig. 3{\em b} shows the power spectrum if
the sinusoid with frequency $f_1$ is subtracted from the
original data set.  There are still one-day aliases present
in the power spectra, owing to the fact that almost all the
data were taken in Arizona and New Mexico, which are almost
at the same longitude.  But Figs. 3{\em a} and 3{\em b} indicate that 9 Aur's
photometric variations can be described well by two sinusoids.

How can we decide if $f_1$ or $1 - f_1$ is the true
frequency? Firstly, $f_1$ is the highest peak in Figs. 2 and 3{\em a}.
That is good, but not definitive proof.  In Fig. 4 we
show the data from a particular 32 day period in the fall of
1993, in which we have folded the data by $P_1$
= 1.25826 d after subtracting out the sinusoid
characterized by $P_2 = 1/f_2 = 2.89282$ d.  Note the two sets of 4 points
obtained on Julian Dates 2449297 and 2449312, respectively.  If we
had folded the data by 1/(1 - $f_1$) $\approx$ 4.87 days, the points
from these two nights would appear as nearly vertical ``posts''
in the folded light curve instead of following the general
run of data from other nights.

 \begin{figure}
 \vspace{70mm}
 \caption{Differential V-band photometry of 9 Aur vs. BS 1561 from
  Julian Dates 2449297.7 to 2449328.9, folded by $P_1 = 1.25826$ d,
  after subtraction of the sinusoid with $P_2 = 2.89282$ d.
  The open triangles represent data from JD 2449297, while the open circles
  represent data from JD 2449312. $\Delta V$ is in the sense 9 Aur
  {\em minus} BS 1561.}
 \end{figure}

A further piece of evidence that $f_1$ and not 1 - $f_1$
is the true frequency comes from an analysis of the data
obtained by Guinan and McCook from 1993 January 30 to February 26.  We
obtain $\phi_1 = -0.143 \pm 0.011$ using the same epoch,
which compares very well with $\phi_1 = -0.114 \pm 0.016$
from Table 2.  A folded plot of the early 1993 data is to be
found in Krisciunas (1994) and is not reproduced here.

A similar analysis gives $\phi_2 = -0.419 \pm 0.026$ for
the early 1993 data, vs. $\phi_2 = -0.481 \pm 0.018$ from
Table 2.  These differences are not significant, or indicate a
value for the frequency $f_2$ slightly different from the one
adopted (see below).

This is strong evidence not only that $f_1$ and $f_2$
are true frequencies, but that the physical mechanism
associated with the photometric variations can be stable over
time scales of one year or longer.  Interestingly,  $f_2$ but
{\em not} $f_1$ shows up in the power spectrum of radial
velocities (see below).  One unresolved issue is the
variations of the amplitudes of the sinusoids.  We find that
$A_1$ in particular varies over quite a range (see Fig. 5),
making the prediction of future variations of the star
impossible, unless it can be shown how the amplitudes vary with time.

 \begin{figure}
 \vspace{70mm}
 \caption{Amplitudes $A_1$ and $A_2$, in mmag, vs. time.  The
  error bars for the abscissa represent the range of dates for the
  subsets of the data.  The error bars for the ordinate are derived from
  the least-squares Fourier fit.  The solid dots connected by a dashed
  line are for $A_1$, while the open circles connected by a solid line
  are for $A_2$. The right-most points have been slightly offset
  from each other in the X-direction for display purposes.}
 \end{figure}

 \begin{figure}
 \vspace{70mm}
 \caption{Differential B-V colors for 9 Aur vs. BS 1561.  The data cover
  the same range of dates, and symbols are the same as in Fig. 4.}
 \end{figure}

Since Guinan and McCook obtained equivalent B-band data, we can
investigate the variations of color in 9 Aur.  In Fig. 6 we
show the folded plot of $\Delta (B-V)$ colors from data of the
same period covered in Fig. 4.  Since ``$\Delta$'' is in the sense
9 Aur {\em minus} BS 1561 and the least positive $\Delta (B-V)$ corresponds
to the bluest color for 9 Aur, one can clearly see that
9 Aur is bluest (i.e. hottest) when it is brightest.  This is the case
for both the $f_1$ and $f_2$ sinusoids, since the phases $\phi_i$ derived
from the $B-V$ colors match the phases derived from the V-band photometry
(compare Tables 2 and 3).  We note that the mean B-V amplitudes of the
1993/4 season about about one-third of the V-band amplitudes.

\begin{table}
\caption{Amplitudes ($A_i$, in mmag) and phases ($\phi_i$) of Fourier fit
to the 1993/4 B-V colors using epoch JD 2449000.0 and frequencies
$f_1 = 0.79475$ and $f_2 = 0.345684$ d$^{-1}$.  The standard deviation
of an individual data point in the two frequency fit is $\pm$ 5.3 mmag.
Note that, within the errors, the phases match those of the V-band
photometry given in Table 2.}
\begin{tabular}{crr}\hline
$i$ & \multicolumn{1}{c}{$A_i$} &  \multicolumn{1}{c}{$\phi_i$} \\
\hline
  $1$  &  $4.9 \pm 0.7$ & $-0.106 \pm 0.023$ \\
  $2$  &  $3.7 \pm 0.7$ & $-0.462 \pm 0.030$ \\
\hline
\end{tabular}
\end{table}

 \begin{figure*}
 \vspace*{110mm}
 \caption{Radial velocities of 9 Aur, obtained by Griffin at Haute Provence
  Observatory.  Top---data of JD 2449347.0 to 2449353.0; middle---data of
  JD 2449353.0 to 2449359.0; bottom---data of JD 2449359.0 to 2449365.0.
  The middle and bottom panels have been vertically offset by -12 and -24
  km s$^{-1}$, respectively.}
 \end{figure*}

\section{Radial Velocities}

A number of reliable values in the literature (Abt \& Levy 1974; Takeda 1984;
Duquennoy, Mayor \& Halbwachs 1991) suggested that the radial velocity
of 9 Aur was slightly variable.
After Griffin demonstrated in early 1993 that the radial velocity of 9 Aur
was indeed variable, a concentrated set of
83 data points was obtained from 1993 December 25 to 1994 January 9.  These
are shown in Fig. 7.  Five more values were obtained from 1994 February
16 to 20, and seven more from April 29 to May 4.  The radial velocities
range from $-5.50 \pm 0.61$ to $+5.15 \pm 0.66$ km s$^{-1}$.  One can
clearly see hour to hour and day to day variations.

In Fig. 8 we plot the power spectrum of the radial velocity data shown in
Fig. 7.  $f_2$ is clearly the dominant peak in the power spectrum.
Interestingly enough, while the frequency $f_1$ and its one day alias
$1 - f_1$ are clearly present in the photometry, $f_1$ apparently does not
show up in the radial velocity data.  In Fig. 8 there is a peak {\em near}
$1 - f_1 \approx$ 0.20 d$^{-1}$ (namely at 0.248 d$^{-1}$).  The small peak
{\em near} $f_1$ is not significant.   We do not understand
how {\em only} the one day alias of $f_1$ should appear, but
not the frequency itself, if indeed the peak near $1 - f_1$ is the one day
alias of $f_1$.)  It could be that during the time
most of the radial velocities were obtained (a period of only
15 days) the frequency $f_1$ was in abeyance.  Future
observations are clearly warranted.

Since high degree spherical harmonics delineate a large number of regions
on the star (which are alternatingly moving in and out or transversely),
while low degree harmonics delineate a small number regions (e.g. 2 to
4 for $\ell$ = 1 or 2), one would only see radial velocity variations in
a star pulsating radially or in a low order harmonic.
It follows that $f_2$ must be related to a low degree harmonic.

 \begin{figure}
 \vspace{70mm}
 \caption{Power spectrum of radial velocities shown in Fig. 7. The
  frequency $f_2$ and its one day alias 1 + $f_2$ are indicated.}
 \end{figure}

Since $f_1$ apparently does not show up in the radial velocity data, it
seems unwise to make a two frequency fit to the radial velocity data to
derive the phases and amplitudes.  The peak of $f_2 \approx 0.36 \pm 0.03$
is close to $f_2 = 0.345684$ from the photometry.  Adopting
the latter value and using all 95 radial velocities available from the
1993/4 season, we find $A_{RV} = 2.00 \pm 0.27$ km s$^{-1}$ and
$\phi_{RV} = -0.139 \pm 0.021$ (which is the phase of maximum {\em positive}
radial velocity).  The standard deviation of an individual
point in this single sinusoid fit
is $\pm$ 1.81 km s$^{-1}$, about three times the typical
internal error of a Coravel radial velocity of this star.  Carrying out a
two sinusoid fit to the data does not improve this standard deviation
significantly.

The next question to ask is: Are the
radial velocities in phase with the photometry?  In Fig. 9 we show all 95
radial velocities from the 1993/4 season folded by $P_2$,
using the epoch of zero phase derived from the photometry of the same season.
The data qualitatively suggest
that 9 Aur has the most negative radial velocity (i.e. most
rapid motion {\em towards} us) close to the time of minimum light for the
$f_2$ sinusoid.  The observed phase lag is $\Delta\phi = 0.158 \pm 0.028$
in the sense that the most negative radial velocity occurs after minimum
brightness.  This phase lag is statisticaly significant, but given the
scatter of the data in Fig. 9, it is probably unwise to make much of it.
Balona \& Stobie (1979) and Watson (1988) discuss how the amplitudes
and phases of the V-band photometry, photometric colors, and
radial velocities can be used to investigate the pulsational mode(s) of
a star, but given the small range of the 9 Aur data compared to the errors,
we feel that detailed modelling along these lines is beyond the scope of
this paper.

 \begin{figure}
 \vspace{70mm}
 \caption{Folded plot of radial velocity data from the 1993/4 season,
  using epoch of zero phase and period derived from the photometry.  The
  sinusoid shown is derived from the Fourier fit of the radial velocities.
  Qualitatively, the star's maximum rate of expansion corresponds to the
  time of minimum light for the $f_2$ sinusoid.}
 \end{figure}

Another question to ask is: Do the radial velocities of
the 1993/4 season phase up with the radial velocities from
the previous season?  This is difficult to say with the same
certainty as can be said from the analysis of the photometry,
since there were only 21 radial velocities obtained on 13
nights in 1993 February and March.  An analysis of the early
1993 radial velocities shows that $f_2$ is once again the
biggest peak in the power spectrum, and we obtain $\phi_2 = -
0.060 \pm 0.043$ for that data set.  (The data set is not
substantial enough to say anything certain about the presence
or absence of $f_1$.)

\section{Revision of the periods}

The data of the most recent season give us accurate working
periods, which allow us to investigate if data sets from more than one
year give nearly equal phases.  It turns out that this is the most
effective way to adjust the periods so that recent data may be compared
with data obtained in the future.

{}From Griffin's radial velocity data of the
past two seasons we find $\Delta \phi_2 = -0.079$
over 125 cycles of 2.89282 days, or an error in the period of
0.00183 d.  From the photometry we find $\Delta \phi_2 = -
0.062$ over 97 cycles, or an error in the period of 0.00185
d.  We obtain a revised period $P_2$ = 2.89282 + 0.00184 =
2.89466 d, or $f_2 = 0.345464$ d$^{-1}$.

\begin{table}
\caption{Observed phases $\phi_1$ for three consecutive seasons
of photometry of 9 Aur, using epoch JD 24449000.0 and assuming
$f_1$ = 0.79475 d$^{-1}$.}
\begin{tabular}{rrr}\hline
$Season$ & $Mean \hspace{1 mm} JD$ &  \multicolumn{1}{c}{$\phi_1$} \\
\hline
1991/2 & 2448657.98 & $-0.205 \pm 0.028$ \\
1992/3 & 2449031.15 & $-0.143 \pm 0.011$ \\
1993/4 & 2449312.31 & $-0.114 \pm 0.016$ \\
\hline
\end{tabular}
\end{table}

In Table 4 we show the corresponding phases of the $f_1$ sinusoid
for the past three seasons of photometry.
When plotted vs. the mean Julian Date of the observations in question they
give a tight linear fit.  We may correspondingly adjust the period
$P_1$ = 1.25826 -- 0.00022 = 1.25804 d, or $f_1$ = 0.79489 d$^{-1}$.

Subsequent photometry or radial velocities would be needed to confirm
if we have really determined $P_2$ and $P_1$ as accurately as they are
given in this paper.  To do so would be significant, for it
would mean that one or both of the frequencies revealed in
the data are indeed very stable.  Knowing the periods exactly is not
important for the purposes of this paper.  What is important is:
1) there are two periods (not one or 100); 2) the periods are not closely
spaced (as in $\gamma$ Dor); and 3) one period is not a simple
multiple of the other ($P_2 / P_1 \approx 2.30$).

\section{Line widths and fractional line depths}

 \begin{figure*}
 \vspace*{200mm}
 \caption{Coravel autocorrelation diagrams of 9 Aur data from 1993/4
  season.  They are labeled with the corresponding Julian Dates (minus
  2440000).  The line fits are given by the Coravel data reduction
  software.  Two sets of data which had poor baseline fits on the positive
  velocity end have been eliminated from the analysis of this paper.}
 \end{figure*}

\setcounter{figure}{9}

 \begin{figure*}
 \vspace*{200mm}
 \caption{Continued}
 \end{figure*}

Coravel works by scanning the star spectrum back and forth across a mask which
is analogous to a high-contrast photographic negative of the spectrum.  The
data
acquisition system produces a profile that represents the cross-correlation
between the spectrum and the mask; it can be thought of as the mean profile
of the stellar lines.   The reduction procedures determine the radial velocity
from each cross-correlation profile, and also give the line depth and a
parameter characterizing the width.   The width parameter can be interpreted in
terms of the projected rotational velocity {\it v} sin {\it i} of the star
(Benz \& Mayor 1981).

In the case of 9 Aur the mean projected rotational velocity is
17.80 km s$^{-1}$ from all the Coravel data, 1992 April 23
to 1994 May 4.  The formal internal error of {\it v} sin {\it i} is
$\pm 0.33$ km s$^{-1}$, but a more realistic uncertainty is $\pm 1.0$
km s$^{-1}$.  This implies a rotational period of 4.66 days times sin {\it i},
adopting a radius of 1.64 $R_{\odot}$ from Mantegazza {\em et al.} (1994).
However, since the non-radial oscillations that we assert to be occurring on
9 Aur are broadening the line profiles, perhaps we should consider that the
projected rotational velocity is no more, and may be less, than corresponds to
the smallest observed values of the line width ($\approx$ 15 km s$^{-1}$),
in which case the true rotational period of the star is not less than about
5.5 days times sin {\it i}.

In Fig. 10 we show the 95 stacked autocorrelation
diagrams from the 1993/4 season.  There are different numbers
of line profiles per graph because we wanted to avoid
splitting up data from a given night.  But it is obvious that
the line width and fractional line depth of 9 Aur is not
constant.  This is a key signature of a star pulsating in one
way or another.

For comparison we investigated the variance of the line
profiles and fractional line depths of BS 3325, a star of
spectral type similar to 9 Aur.  The observations were made
on many of the same nights, with the instrument in the same
configuration.  (BS 3325 is a spectroscopic binary whose radial
velocity ranges 120 km s$^{-1}$ on a time scale of about 4.5 days.
See Griffin, Eitter \& Appleton (1995) for details.)
In the case of BS 3325 the
line widths and fractional line depths produce classically
Gaussian and much narrower histograms (a factor of 2 narrower)
compared to 9 Aur.  Thus, we can be assured that the observed
variations of line profile, line width, and fractional line
depth observed in 9 Aur are inherent in 9 Aur and are not
artefacts of the Coravel instrument or data reduction procedures.

 \begin{figure}
 \vspace{70mm}
 \caption{Power spectrum of the line widths of 9 Aur from the Coravel
   data of the 1993/4 season.}
 \end{figure}

 \begin{figure}
 \vspace{70mm}
 \caption{Power spectrum of the fractional line depths of 9 Aur from the
   Coravel data of the 1993/4 season.  The frequency labeled X could
   be equal to 1 - $f_2$ (a one day alias).}
 \end{figure}

In Figs. 11 and 12 we show the power spectra of the line
widths and fractional line depths derived from the Coravel
data.  Compared to the power spectrum of the radial
velocities shown in Fig. 9, frequency $f_1$ is clearly
present in the power spectra of line widths and fractional
line depths.  Since line profile variations are much more
pronounced in higher degree spherical harmonics ($\ell \geq 3$
-- see Vogt \& Penrod 1983), at face value the evidence
presented here points to the following conclusion: in 9 Aur
$f_2$ comes from a low degree harmonic ($\ell$ = 1 or 2) while
$f_1$ comes from a higher degree harmonic.

\section{Discussion}

{}From extensive photometry in 1993 and early 1994 we have
confirmed that there are two primary frequencies present in
the light curve of 9 Aur.  Our best estimate of one frequency
is $f_1$ = 0.79489 d$^{-1}$,
or $P_1$ = 1.25804 d, with an epoch of zero phase (when the
star is faintest) of JD 2449000.20 $\pm$ 0.02.  Our best
estimate for the other frequency is $f_2$ = 0.345464 d$^{-1}$,
or $P_2$ = 2.89466 d, with an epoch of zero phase of JD
2449001.19 $\pm$ 0.05.  Because the amplitudes of these
sinusoids can change, it is not possible to forecast a light
curve.  Future observations might reveal if there is a
pattern to the variations of the amplitudes.

We note that our frequency $f_1$ is essentially equal to
a principal frequency of 0.80 d$^{-1}$ in the light curves of
HD 224638 and HD 224945 (Mantegazza {\em et al.} 1994).  The
two frequencies of $\gamma$ Dor are $f = 1.32098$ and $f =
1.36354$ d$^{-1}$ (Balona {\em et al.} 1994b).  For HD 164615
the principal frequency is $f \approx 1.227$ d$^{-1}$ (Abt {\em et al.}
1983).

The radial velocity of 9 Aur is indeed variable, with an amplitude
(semi-range) of 2.0 km s$^{-1}$.
A power spectrum of the radial velocity data of 1993/4 clearly
shows $f_2$, but the data do not clearly reveal $f_1$.
Should the power
spectrum of the radial velocities of 9 Aur continue to show
only $f_2$ and not $f_1$, it would imply that $f_2$ is
related to a low degree spherical harmonic ($\ell$ = 1 or 2),
while $f_1$ is related to a higher degree harmonic.

We have found that the $B-V$ color of 9 Aur is bluest
when the star is brightest, for both the $f_1$ and $f_2$ sinusoids.
For Cepheids, which pulsate radially, this is also true.  In Cepheids
one observes the maximum
negative radial velocity (i.e. the star is expanding) at the
time of maximum brightness (Joy 1937).  For 9 Aur the maximum negative
radial velocity coincides approximately with the {\em
minimum} brightness of the star, but we can only say this as
it pertains to the frequency $f_2$ because frequency $f_1$
was not obviously present in the radial velocity data.

{}From the absence of certain evidence (i.e. the existence
a close, interacting companion, a lumpy orbiting ring of
dust, or plausible star spot models)
and the demonstration that the radial velocities, line
profiles, line widths, and fractional line depths of 9 Aur
are variable with the very same frequencies found in the
photometry, we believe that 9 Aur is exhibiting non-radial
pulsations.  They must be non-radial gravity modes because
the periods of variation are at least an order or magnitude
slower than the fundamental radial pulsation period.

Eddington once said that observations should not be
believed unless they are supported by a good theory (Haramundanis
1984).  This highlights one of the problems of this research topic.  So
far no successful pulsation mechanisms have been discovered
which result in realistic models of F dwarf stars exhibiting
non-radial gravity modes (A. Gautschy, private communication).
Applying canonical thermal time scale arguments in connection
with the working of a $\kappa$-mechanism, one would expect a large
number of modes to be excited simultaneously.  (The involved thermal time
scale is of the order of a day and the period spacing around an oscillation
period of one day is $3 \times 10^{-3}$ days.)  The observed situation
calls for a sharp resonance condition that destabilizes particular
modes only.  Since it seems quite certain that we have
serendipitously discovered a new class of variable stars, it
is important to produce better models of early F stars so
that we can understand under what conditions these stars
show evidence for non-radial pulsations.

The most useful data to obtain in the near future would
be near-simultaneous UBV or $uvby$ photometry and high resolution
spectroscopy of a number of the unusual F stars -- data
obtained at multiple sites around the globe.
Another very useful observing program would be to
follow up the type of observations carried out by Antonello
\& Mantegazza (1986).  Array photometry of open clusters, if
obtained to the 0.01 mag level, should reveal new F dwarf
variable stars with ``slow'' periods such as the stars
discussed here.  This is a big undertaking, because these
stars have periods on the order of one day, so either massive
quantities of data must be obtained at a single site, or data
must be obtained at multiple sites to derive the true periods
of the variations.

F dwarf stars are only slightly more massive, slightly
larger, and slightly hotter than the Sun.  It is suprising
that a number of them found outside the cool edge of the Cepheid
instability strip in the HR Diagram have been found to be variable at the
0.10 mag level, whereas the Sun is constant to better than
0.001 mag (Hudson 1988).  Our slowly varying F stars are definitely {\em not}
$\delta$ Scuti stars, and they remind us that new phenomena can still be
discovered in otherwise normal stars.

\section*{ACKNOWLEDGEMENTS}

    RFG is most grateful to Dr. M. Mayor and the Observatoire de
$\rm Gen\grave{e}ve$ for
the provision of observing time on the 1-m Geneva telescope and Coravel radial
velocity spectrometer at Haute Provence, and to Dr. A. Duquennoy for retrieving
the relevant radial velocity observations from the Coravel data base and for
performing the reductions.   The costs of the six observing trips
were defrayed by the UK Science and Engineering Research Council.

    KK thanks J. Matthews, E. Poretti, C. Waelkens, A. Gautschy, and,
L. Balona for many useful discussions.  We also thank the referee, D.
Kurtz, for useful comments and references to other work.

    Some of the photometry presented in this paper was obtained with the
Automatic Photoelectric Telescope of the Four College Consortium (FCC).
The FCC consists of The College of Charleston, The Citadel, Villanova
University, and the University of Nevada at Las Vegas.  Funding is provided
by the National Science Foundation (NSF) grant AST-86-16362.

    During the writing of this paper we were saddened to learn of the
death of A. Duquennoy in an automobile accident.  Without his recent efforts
this paper would have been indefinitely delayed.

\bsp

\end{document}